\documentstyle[epsf,psfig]{aipproc}

\newcommand{\be}{\begin{equation}}
\newcommand{\ee}{\end{equation}}

\begin{document}

\title{On Financial Markets Trading}

\author{Lorenzo Matassini and Fabio Franci}

\address{Max-Planck-Institut f\"ur Physik komplexer Systeme\\
         N\"othnitzer Str.\ 38, D 01187 Dresden, Germany\\
		 Tel: +49 351 8711803 \hspace{1mm} Fax: +49 351 8711999\\
         email: lorenzo@mpipks-dresden.mpg.de}

\maketitle

\begin{abstract}

Starting from the observation of the real trading activity, we propose a model of a stockmarket simulating all
the typical phases taking place in a stock exchange. We show that there is no need of several classes of agents
once one has introduced realistic constraints in order to confine money, time, gain and loss within an appropriate
range. The main ingredients are local and global coupling, randomness, Zipf distribution of resources and price 
formation when inserting an order. The simulation starts with the initial public offer and comprises the
broadcasting of news/advertisements and the building of the book, where all the selling and buying orders are
stored. The model is able to reproduce fat tails and clustered volatility, the two most significant characteristics
of a real stockmarket, being driven by very intuitive parameters.

\vspace*{4mm}

\vspace*{4mm}

\end{abstract}

\section*{Introduction}

The dynamics of the stock market oscillations is still object of a great debate. Variation of stock prices is usually
considered a random process and there are many alternatives about the proper model of the distribution of return.
In any case, some universal features have been found: They resemble the scaling laws characterizing physical systems 
dominated by the interaction of a large number of units.
It is therefore of a great interest the introduction of a model that is able to reproduce these aspects through a
proper tuning of its parameters. Also interesting is the understanding of the direct influence of them, provided they
have a physical meaning.

Previous models are all essentially based on the assumption that two different kinds of economic agents are interacting
in the market: Some authors \cite{bischi} call them \emph{dealers} and \emph{savers}, others \cite{lux_nature} use the
names \emph{fundamentalists} and \emph{noise traders} (further distinguishing between optimistic and pessimistic), others
more \cite{bak} say \emph{rationals} and \emph{chartists}. If one goes into the details of these models, one sees that
\emph{fundamentalists} follow the premise of the efficient market hypothesis in that they expect the price
to follow the fundamental value of the asset. A fundamentalist trading strategy consists of buying when the
actual market price is believed to be below the fundamental value and selling in the opposite case.
\emph{Noise traders}, on the other hand, do not believe in an immediate tendency of the price to revert to its underlying 
fundamental value: They try to identify price trends and consider the behavior of other traders as a source of information, 
giving rise to the tendency towards \emph{herding}.

Looking at the scaling law characterizing physical systems where large numbers of units interact
\cite{haken,galluccio,mant_nature,mantegna}, anyway, one observes that there is no need to introduce different classes of agents.
Since the details of the circumstances governing the expectations and decisions of all the individuals are unknown to
the modeler, the behavior of a large number of heterogeneous agents may best be formalized using a probabilistic
setting. Such a statistical modeling concept has a certain tradition in the so-called \emph{synergetics} literature
\cite{haken} which has adopted techniques from elementary particle physics to study various problems of social 
interactions among humans \cite{ramsey}. In a statistical approach the properties of macro variables are not necessarily 
identical to those of the corresponding micro variables nor does the mere aggregation of micro components always yield 
sensible macroeconomics relationships.

The most important aspect to be taken into account, therefore, is the behavior of the single particle and the
coupling between units. A realistic model, in our opinion, should be able to avoid
the distinctions among classes of traders, because they introduce some collateral problems to be solved.
In \cite{lux_nature}, for instance, it is necessary to let people move from one group to another and to introduce an 
exogenous change of the fundamental value, but these two requirements sound somehow artificial. It is in fact clear
that without the fundamentalists the price would follow just one trend at infinitum and without the noise traders
the price would never escape the range of its fundamental value. In \cite{bak} this fact becomes clear, since already 
a 20\% of fundamentalists is enough to confine the market price within the range of the rational traders. Just very short
deviations occur before common sense was restored.
Some authors even assume that the relative changes of the fundamental value are Gaussian variables, leading to the
completely unrealistic situation where rational traders do invest according to some source of randomness.

In \cite{lux_nature} the authors want to show that although the \emph{news arrival process} lacks both power-law scaling 
and any temporal dependence in volatility \cite{yousse}, their model generates such a behavior as the result of the 
interactions between agents. But then one realizes that the market price follows very closely the fundamental value,
i.e. it is itself almost random. In \cite{lux2}, furthermore, it has been shown that behavioral models can produce a 
shape of the \emph{return distribution} remarkably similar to that derived from empirical data; therefore any
unobservable news arrival process is not necessary in order to explain the salient characteristics of empirical 
distributions of returns. On the contrary, the high peaks and fat tails property can be derived from the working of the 
market process itself.

In our model there is only one type of investor, whose goal is maximizing the profit while minimizing the risk. Every trader
has a limited amount of money and a given disposition towards investment. Furthermore the time comes into account, since
a given gain has a different meaning whether realized within few weeks or after several years. At the beginning of the
simulation, one agent is supposed to be the central bank responsible of the \emph{Initial Public Offer} ({\bf IPO}):
All the shares belong to one agent. We provide a mechanism to generate news and advertisements as a way to introduce
a global coupling into our model. During the IPO, traders feel a strong pressure to buy and in fact almost all of them 
want to order some quotas. Since the bank responsible of the IPO cannot buy back any shares at the moment, this
transient has a limited length and the simulation enters the permanent regime after few iterations.
The main building blocks of the model are the price formation and the book, where all the pending orders are stored.
Every trader, when willing to buy a share, has to identify a fair price according to past values of the price, opinion
of the media and suggestions coming from acquaintances. This is the price with which he/she would like to enter the
market; starting from it every agent keeps in mind a target price and a stop-loss price (respectively according to
the desired gain and the maximum loss). 
They are of fundamental importance to decide whether and when to sell some shares, together with a threshold in time.
We neither introduce any fundamental price nor make use of any exogenous input.
In the following sections we describe the model in more details, present some simulation runs and study the influence
of the parameters.

\section*{The model} 

We consider $N$ traders who want to trade a stock with $M$ shares on a common market. For simplicity only one stock is
involved in the model. The information related to every trader are the following:

\begin{itemize}

\item Initial amount of money: According to \cite{zipf,marsili,zipf2,takayasu}, it is likely that the Zipf's law
      is quite universal. On the light of this observation, we have decided to distribute the money to traders following
	  such a distribution. A lot of money to a small amount of agents, few resources to the majority of people.

\item Inclination towards investment: Usually traders tend to keep cash a part of their resources, in order to be
      able to have money to exploit the market at special time.

\item Number of shares owned.

\item List of friends with which sharing information. Through this local coupling we are able to model the herding
      behavior. In \cite{marsili} it has been shown that interactions of order higher than pairwise are not relevant
	  in the dynamics that lead to Zipf's law.

\item Invested money, to keep trace of the average buying price for further purchase.

\item Cash: Amount of resources immediately available, sometimes called liquid.

\item Expected gain, randomly chosen between $g_{min}$ and $g_{max}$. Once such a profit has been realised the
      trader wants to sell the shares in order to convert this invested money into cash.

\item Maximum loss, randomly chosen between $l_{min}$ and $l_{max}$. It is sometimes better to sell the shares losing
      some money than keeping them during a crash (and therefore losing more money after that).

\item Personal target price, decided when inserting the buying order.

\item Personal stop loss, decided when inserting the buying order.

\item Threshold: Amount of time after which the trader may start to change idea about the investment: In real situation
      the gain alone is not enough to say whether an investment has been a good one or not. Everyting has to be related
	  with time (10\% gain in one year is of course a worst investment than 5\% gain in one month if we do not consider
	  speculation taxes).

\end{itemize}

When an agent wants to trade, a new record in the book is created. All the orders are stored according to the type 
(buy or sell), to the price and to the submission time. A transaction occur whenever the cheapest price among the sell 
list matches with the most expensive offer in the other list: That price defines the market price of the stock at that 
particular instant of time (tick). In the following table we can see the first five levels of the book. Every line 
contains information about time, trader, number of involved shares and desired price. Entries are ordered according 
to the price (and eventually to the time for equal prices), in a decreasing way for the buying list and in an increasing 
manner for the selling list.

\vspace*{1cm}
\centerline{\psfig{file=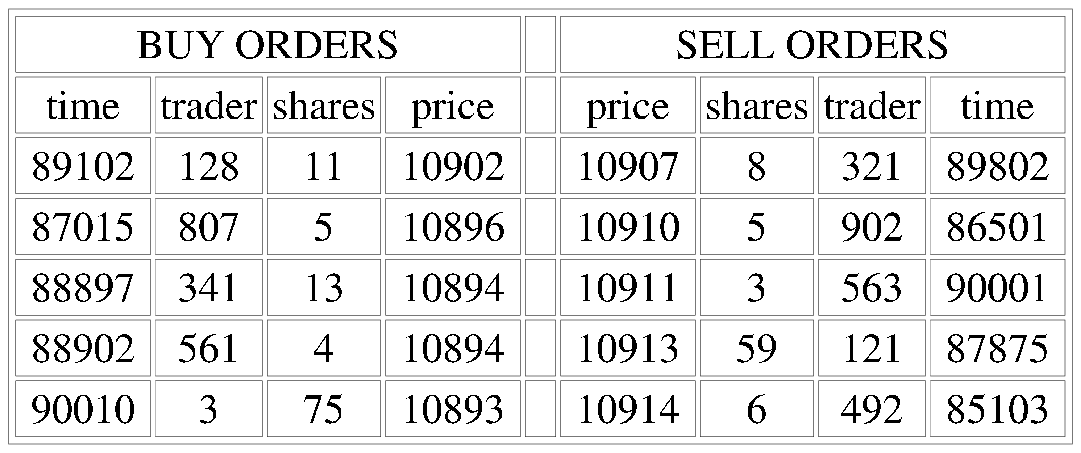,width=11cm}}
\vspace*{0.3cm}
{\noindent \small {\bf TABLE 1.} Example of the first five levels of the book. No transaction can take place because the 
highest buying price is smaller than the cheapest selling order. Entries are ordered according to price and occurrence 
time.}
\vspace*{0.7cm}

Once terminated the IPO phase, a typical simulation iteration involves the following steps:

\begin{itemize}

\item Selection of trader: This is done in a completely random way, without taking into account the different weight
      of agents, nor any other characteristic. This part represents material for a future work, where we can introduce
	  some feedback between the type of trader and how often he/she trades.

\item Control of the presence of some pending order: The trader has NO pending order.
	  
	  \begin{itemize}
	  
	  \item Check whether the trader owns shares. If the trader owns NO share, then:
	  
	        \begin{itemize}
			
			\item Evaluate the probability of a buying order for the trader. Formulate the fair price according to
			      Eq.\ref{eq:prezzo} and target (stop loss) price according to expected gain (maximum loss).
			     
			\end{itemize}
			
	  \item Else (the trader has one pending order):
	  
	        \begin{itemize}
			
			\item Check the market price and since when the trader owns the shares. If the actual price is
			      below the stop loss then evaluate the probability that the trader performs a caution sell and
				  the probability related to an averaging purchase.
				  This decision is taken according to the news suggestions (buy, keep or sell) and to the  
				  available amount of money.
				  
				  If the market price is over the target then the trader will sell all the shares, inserting
				  a limit order with a price in the range (target price, market price).
				  
				  If the price is in the range (stop-loss price, target price) and the trader has bought the
				  shares a long time ago (longer than the threshold), evaluate the probability to sell,
				  inserting a limit order at the previously computed target price. 
			
			\end{itemize}
	  
	  \end{itemize}
	  
\item Else (for simplicity we suppose that every trader cannot afford more than one pending order simultaneusly):
	  
	  \begin{itemize}
	  
	  \item Check the kind of the order. If it is a request to SELL, then:
	  
	        \begin{itemize}
			
			\item If the order is sufficiently old (older than the threshold), then evaluate the probability that 
			      the trader decides to remove it.
			
			\item If the market price is below the stop loss price, then evaluate the probability that the trader 
			      performs a market sell or an averaging buy.
			
			\end{itemize}
			
	  \item Else (request to BUY):
	  
	        \begin{itemize}
			
			\item Evaluate the probability that the trader decides to convert the limit order in a market order.
			      This is done according to the media and the acquaintances.
				  
		    \item Compare the time with the threshold: If the order is too old, then evaluate the probability that 
			      the trader removes the order. It is also possible to change just some parameters of the order.	  
			
			\end{itemize} 
	  
	  \end{itemize}	 
	  
\item Check the book. If there is a price-matching (the cheapest selling order equals the highest buying price), then:

      \begin{itemize}
	  
	  \item Define this price as the market price at the actual time (tick).
	  
	  \item If the buyer has got all the desired shares then this buying order is removed from the list and
	        the trader can immediately place a selling order (through an evaluation of the probability).
			
	  \item If the seller has been able to sell all the desired shares, then this selling order is removed from the
	        corresponding list.
	  
	  \end{itemize}

\end{itemize}

When formulating the fair price, every trader makes the decision according to the opinion of some acquaintance 
(contained in the price $p_1$), to the media (they suggest the price $p_2$) and to some past values of the stock price 
itself (through the forecasted price $p_3$, inspired by \cite{baptista}).

\begin{itemize}

\item $p_1$ represents the opinion of the friends of the selected trader. Every agents who owns shares, has his/her own
target price. We average all the target price of the friends which are making profit (who would believe to people which are
losing money?) and then establish an \emph{enter price} rescaling it by the expected gain ($eg$) of the selected 
trader:
$
p_1= \frac{1}{n} \sum_{i=1}^n {tp_i} (1-eg)
$

\item $p_2$ represents the target price broadcasted by the media, in our model an internal global information of the
system\footnote{This behavior is realistic because in real life these infos are still internal, since also media and 
the big istitutions invest in the stock market and often broadcast biased news
in order to make easy money.}. 
It is generated starting from the last price $p_u$, comparing the current ratio between the number of buyers and the 
number of sellers in the book ($b_u$ stands for book unbalance) with the parameter $B$ (unbalance of the book), in 
the following way:
$p_2=p_u (1+r_{max})$ if $b_u>B$,
$p_2=p_2$ if $\frac{1}{B}<b_u<B$ or
$p_2=p_u (1-r_{max})$ if $b_u<\frac{1}{B}$.

\item $p_3$ represents the expected price due to the actual trend. It is formulated taking into account the trend over the
past \emph{MEM} prices and comparing it with the last price, $p_3 = p_u+ (p_u-p_{trend})$.

\end{itemize}

Finally, the price $p$ is obtained as a weighted average of $p_1$ $p_2$ and $p_3$ in the following way:

\be
p=\frac{p_1}{\alpha_1}+\frac{p_2}{\alpha_2}+\frac{p_3}{\alpha_3}	
\label{eq:prezzo}
\ee

with $\alpha_1$, $\alpha_2$ and $\alpha_3$ such that $\sum_{i=1}^{3}\frac{1}{\alpha_i} = 1$.

\section*{Simulation Runs}

In every model there is the effort to avoid a parameter explosion: It is in fact clear that, with a sufficiently large
number of degrees of freedom, one can approximate and reproduce every behavior and any feature of a given system.
When modeling, nevertheless, one has to keep in mind that a model is meanungful as long as the parameters have a
clear meaning and a directly observable influence.

\begin{figure}
\centerline{\psfig{file=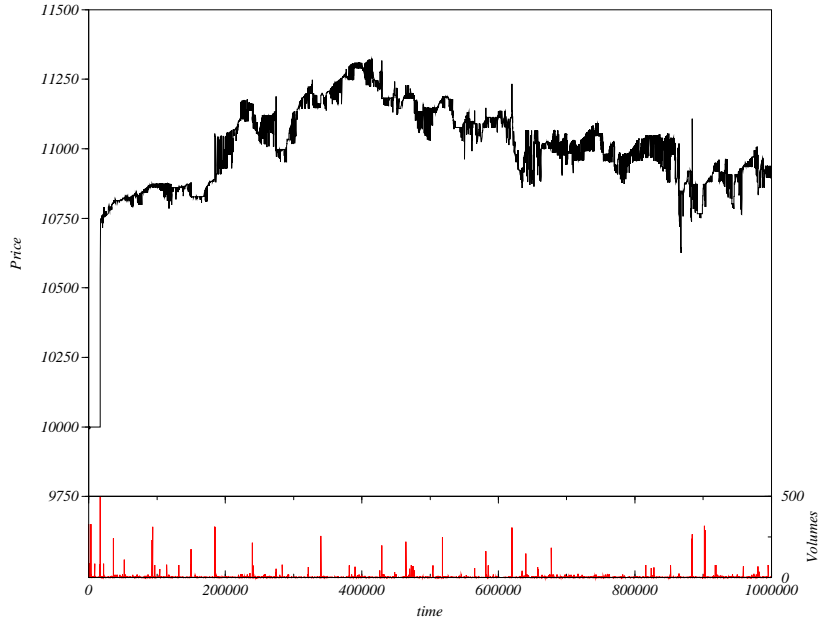,width=11cm,angle=270}}
\caption[]{\small\label{fig:serie}
Typical snapshot from a simulation run. Upper panel: Development with time of the market price.
Lower panel: Development with time of the corresponding volumes of exchange.}
\end{figure}

The result of a typical simulation run is shown in Fig.\ref{fig:serie}: The upper panel is related to the market price, 
whereas the lower panel reports about the corresponding amount of exchanged shares. We can clearly see the IPO phase 
lasting few hundreds of ticks, where the price remains constant because the bank offers the shares to the
traders at a suggested value. After this transient, the pressure made from people without shares becomes visible:
The volumes are high and the price tends to raise. Letting the time evolve, one runs into a tranquil period, with
a slowly oscillating price and very low volumes: Owners do not want to sell because they hope to get more money if 
the wait a little bit, agents without shares want to buy at a lower price.
This dynamical equilibrium is unstable: After a significant increase in the volumes a small burst occurs and the
price increases to a more suitable value for owners. A rally is usually followed by a crash (a kind of settlement), 
as reported in \cite{platt} under the name of \emph{on-off intermittency}, an aperiodic switching between
static, or laminar, behavior and chaotic bursts of oscillations.

We explain in the following the origin, the tuning and the effect of the parameters involved in the model. 

\begin{itemize}

\item Number of traders {\bf N}: The number of active agents trading on the market.

\item Threshold {\bf T}: The number of ticks after which the trader
                         may start to have some doubts about the performed investment. This value
						 ranges among a decade in order to model the different temporal horizons of the investments
						 (differences between short-time speculators and long-time traders).
						 
\item Number of shares {\bf S} and IPO's price {\bf I}: Their product defines how the company is initially worth.
                         The value of these two parameters is chosen after an analysis of the typical IPOs taking
						 place in the european stockmarket.

\item Amount of money {\bf M} initially distributed among the traders. As in the case of the threshold, this is not 
                              equal for all the players. Differently from the threshold, the initial wealth of the
							  traders follows a Zipf's distribution.
							  Inspired by \cite{zipf,marsili,zipf2,takayasu} we have decided 
							  to distribute the richness according to a Zipf's law in the following way: the 20\% of 
							  the traders posses around the 80\% of {\bf M} and among the two groups this rule is 
							  applied again in a recursive way\footnote{As already mentioned in the text, the Zipf's 
							  law is quite universal. Just to provide
							  some examples, the directory size in a hard drive, the CPU time used by processes
							  in a workstation, the number of hits per homepages, the income distribution in
							  companies, the debts of bankrupt companies, the city distributions,
							  they all follow a Zipf's law.}.
							  In this way we are able to model the difference between a normal agent and an
							  istitutional investor and the different effect they produce when they decide to
							  enter the market. Anyway a minimal value of money {\bf m} is provided to all the traders
							  and added to the amount coming from the previous distribution.
							
\item Length of the past values' list {\bf MEM}: Chartists look for trends and patterns in the historical time series
                                                 of the market price. Given the very short correlation in time, it is
												 not so meaningful to look to much in the past, but some differences
												 arise according to the tuning of this value. {\bf MEM} is not
												 constant for all the traders, since some of them have access to more
												 information than others.

\item Unbalance of the book {\bf B}: This value takes into account how balanced are the sell and the buy list
                                          with respect to each other. In case of a strong asymmetry an
										  automatic mechanism generates news and
									      advertisements. Through this parameter we introduce
										  a global coupling in the model.

\end{itemize}

\begin{figure}
\centerline{\psfig{file=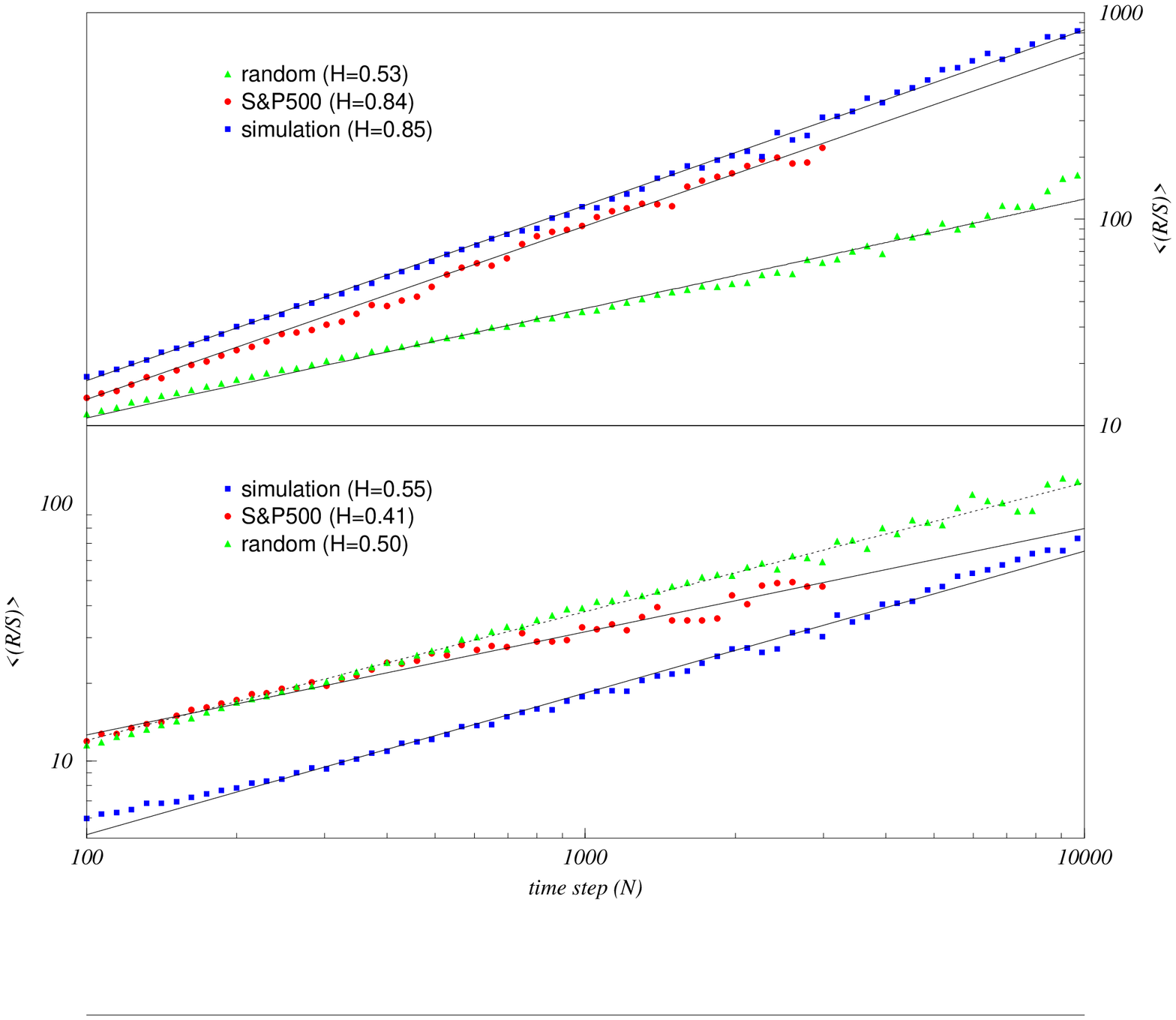,width=11cm,angle=270}}
\caption[]{\small\label{fig:hurst}
Typical snapshot from a simulation run. Estimation of the self-similarity parameter \emph{H}. Upper panel: Absolute
returns. Lower panel: Raw returns. It is interesting to note the emergence of correlations when passing from raw
to absolute returns. This happens to empirical data, a stock belonging to the S\&P500 index, and to our model, but
not to the white-noise process. The strong persistence characterizing the absolute returns of our simulation and
of a real stock is an indication of the so-called \emph{correlated volatility}. This is also visible looking to
Fig.\ref{fig:fattails}, where a clear alternation between tranquil and turbulent periods is present.}
\end{figure}

As already mentioned, universal characteristics exhibited by financial prices comprise a distribution with fat tails
(events with a distance bigger than $3\sigma$ from the average return are not so uncommon as a Gaussian distribution
would forecast) and a correlation in the volatility (alternation between tranquil and turbulent periods). 
In Fig.\ref{fig:hurst} we show the presence of a strong persistence in the volatility. This is done 
estimating the self-similarity parameter H \cite{peng} for raw and absolute returns (being the latters a measure 
of volatility). The lower panel of Fig.\ref{fig:hurst} is related to raw returns: A random time series with the same
variance of our simulated evolution of the price gets a value H=0.50, corresponding to brownian motion. A time series
of a stock belonging to the S\&P500 index gets a sligtly smaller value (H=0.41), with our simulation we have got
H=0.55. The prove of the presence of correlations in the volatility comes from the upper panel, when considering
absolute returns. The random time series is still uncorrelated (H=0.53), while a time series coming from a real or a
simulated stockmarket shows a kind of memory (H=0.84 and H=0.85 respectively).

\begin{figure}
\centerline{\psfig{file=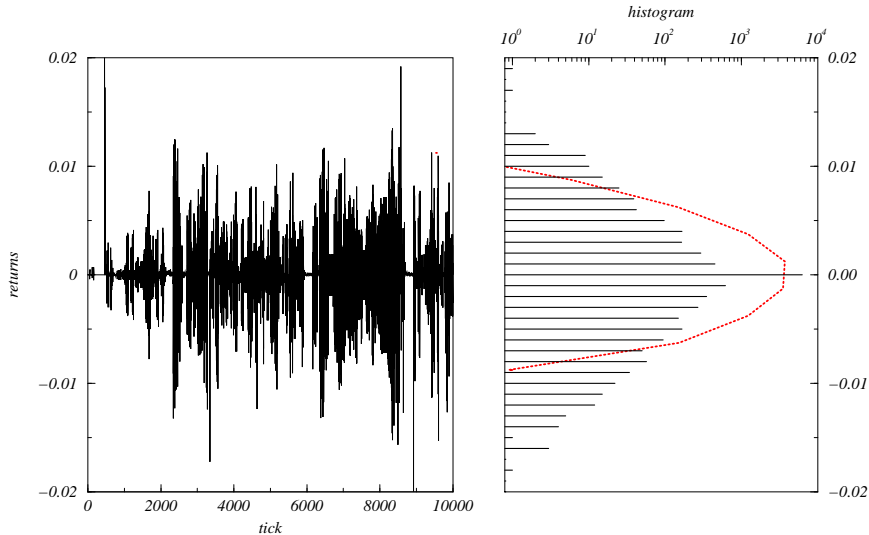,width=12cm,angle=270}}
\caption[]{\small\label{fig:fattails}
Left panel: Price returns. Right panel: Returns distribution. The comparison
with a best-fitted normal distribution (dotted line) reveals the presence of \emph{fat tails}.}
\end{figure}

Fat tails are detectable through the probability density function (PDF) of the returns. In Fig.\ref{fig:fattails} we
show the PDF of our simulated time series together with a gaussian distribution of returns having the same standard
deviation. Extreme events happen with a higher frequency in the simulation, giving raise to the typically observed 
alternation between tranquil period, rally days and crashes.

\section*{Parameters' Analysis}

Let us consider now the following quantity {\bf P}, what we call \emph{buying factor}:

\be
{\bf P} = \frac {{\bf M} + {\bf m} {\bf N}} {{\bf S} {\bf I}}.
\label{fig:P}
\ee

We have obtained significant results only if {\bf P} belongs to the interval $[4,8]\footnote{Just to give some 
colors to these numbers, we can imagine the following situation, not too far away from the reality in a big 
european country. Suppose that 10 million of traders are interested in trading a given new stock and they would 
like to invest, on average, 3000 EUROs. They have an investing power of 30 billions of EUROs.
Suppose now that the new stock comes out with 100 million of shares at an IPO's price of 50 EUROs.
The initial value of the company is 5 billions of EUROs and the value of {\bf P} is 6.}$ and we have the following
explanation. The numerator of Eq.\ref{fig:P} indicates the total amount of money that the traders could invest in 
the stock. The denominator, on the other hand, is an indication of the value of the company at the initial 
public offer. If people do not have enough money (small value of {\bf P}), they are not encouraged to trade and 
the market price decays. If the agents have a too big amount of resources compared to the company's value
(i.e. {\bf P} is big), then the market price starts to increase until the resource limit is reached.

This represents, on our opinion, a realistic upper limit for the market price, since nobody would buy anything if
they cannot afford it (maybe the reality is a little bit different; traders stop buying shares when the price is
too high not because they do not have money, but because they believe that the price is too far away from a
fair price for the given stock, a concept that could be modeled with the introduction of a fundamental value, something
that we strongly want to avoid).

\begin{figure}
\centerline{\psfig{file=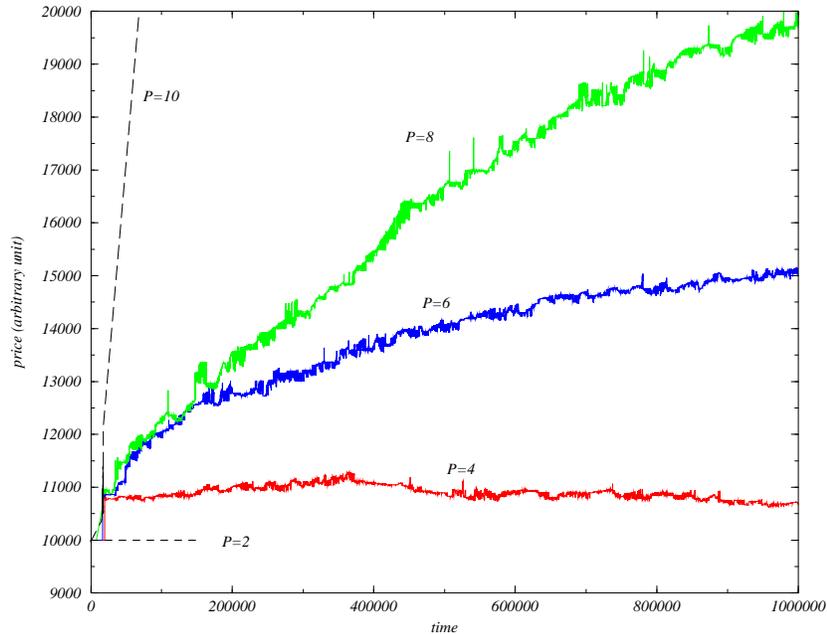,width=11cm,angle=270}}
\caption[]{\small\label{fig:zipf}
Typical snapshot from five simulation runs. Dependence on the buying factor {\bf P}.}
\end{figure}

In Fig.\ref{fig:zipf} we present the dependence on the \emph{buying factor} of a typical simulation run.
Values of {\bf P} in the range $[4,8]$ present the already discussed features. When ${\bf P}=10$ the price increases
very fast, namely it twices after a few thousand of iterations. With ${\bf P}=2$, the model is not able to end
the simulation of the IPO phase: Although traders had enough money to buy all the shares available 
at the bank, they do not want to invest too much at the beginning, according to their \emph{inclination towards 
investment}.

One comment about demand and supply. It had been a common sense in economics for a long time that demand and supply
balances automatically, however, it becomes evident that in reality such balances are hardly be realized for most of
popular commodities in our daily life \cite{verri,takayasu2}. The important point is that demand is essentially a stochastic
variable because human action can never be predicted perfectly, hence the balance of demand and supply should also be
viewed in a probabilistic way. If demand and supply are balanced on average the probability of finding an 
arbitrarily chosen commodity on the shelves of a store should be 1/2, namely about half of the shelves should be
empty. Contrary to this theoretical estimation shelves in any department store or supermarket is nearly always
full of commodities. This clearly demonstrates that supply is much in excess in such stores.

In general the stochastic properties of demand and supply can be well characterized by a phase transition view which
is consisted of two phases: The excess-demand and excess-supply phases. It is a general property of a phase 
transition system that fluctuations are largest at the phase transition point, and this property also holds in 
this demand-supply system. In the case of markets of ordinary commodities, consumers and providers are indipendent
and the averaged supply and demand are generally not equal. The resulting price fluctuactions are generally slow
and small in such market because the system is out of the critical point.

\begin{figure}
\centerline{\psfig{file=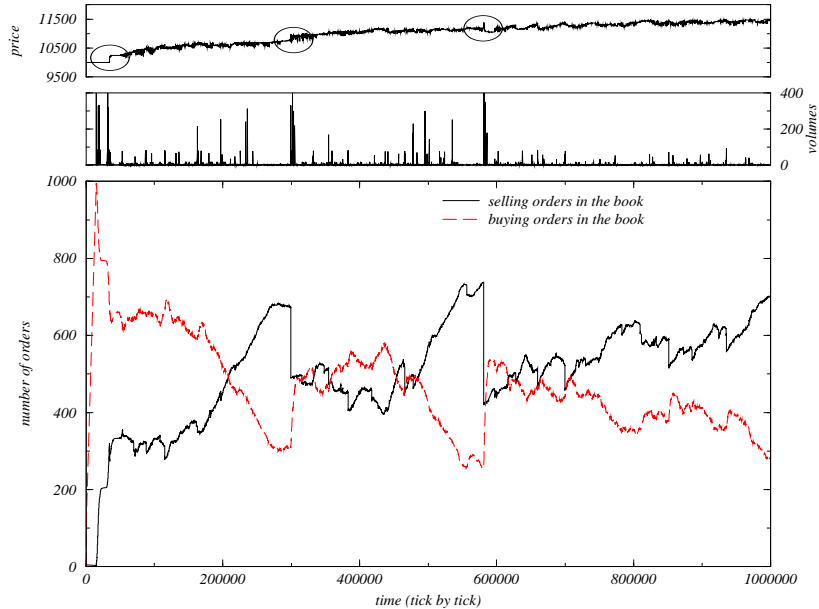,width=11cm,angle=270}}
\caption[]{\small\label{fig:sbilbook}
Upper panel: Market price evolution on time. Middle panel: Volumes of exchanged shares on time.
Lower panel: Evolution of the book. For this simulation we have used N=1000. Almost all the traders have placed
an order and are waiting. Note the symmetry of the two number of orders with respect to the half the number of agents. 
At the beginning the two lists are empty, then almost all the traders want to own some shares and nobody wants to sell,
either because he/she does not own any, or because the price is not convenient. After few thousand of iterations
a dynamical equilibrium is reached.}
\end{figure}

\begin{figure}
\centerline{\psfig{file=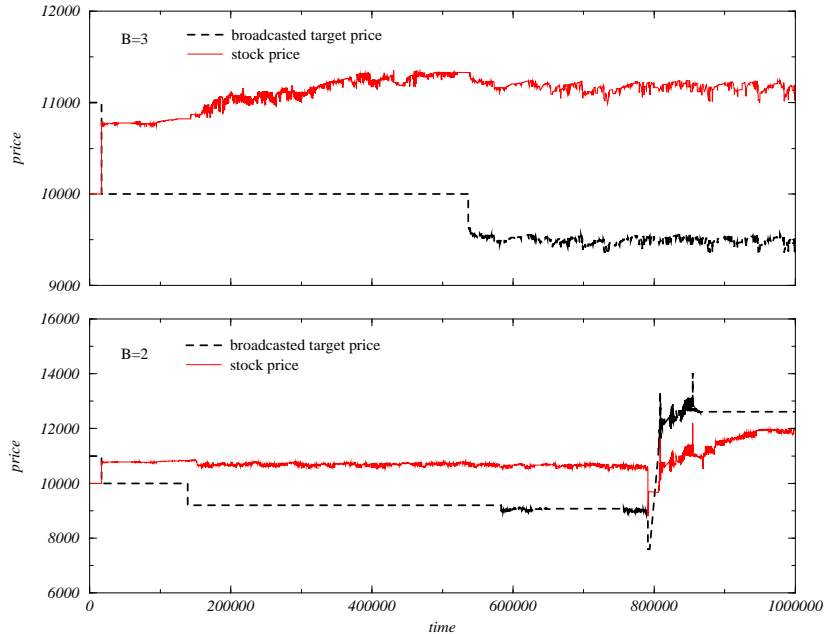,width=11cm,angle=270}}
\caption[]{\small\label{fig:bookunb}
Influence of the parameter B on the evolution of the market price. A smaller value of B (lower panel, B=2)
has a bigger influence on the general opinion, since advertisements are broadcasted more often.}
\end{figure}

On the contrary in an open market of stocks or foreign exchanges, market is governed by speculative dealers who
frequently change their positions between buyers and sellers. It is shown that such speculative actions make demand
and supply balance automatically on average by changing the market price, resembling a kind of \emph{self organized
criticality}. Fig.\ref{fig:sbilbook} is in total
agreement with that: Contrary to \cite{bak} we do not need to impose that the number of the shares has to be half
of the number of traders in order to get a balance between demand and supply. The three circles in the upper panel 
indicate the most extreme events taking place in the price evolution: There is a corresponding movement in the book and
a peak in the volumes. As the system is always at the critical
point the resulting price fluctuations are generally quick and large \cite{takayasu2}. This result is in agreement
with \cite{sornette}, where the authors present an analogy between large stock market crashes and critical points
with log-periodic correction to scaling: Complex systems often reveal more of their structure and organization
in highly stressed situations than in equilibrium. 

In Fig.\ref{fig:bookunb} we report on the influence of the media on
the market price. In particular we note that a small value of B (B=2, lower panel) forces with more strength the 
underlying price than a bigger value of the unbalance parameter (B=3, upper panel). The power of the advertisement!

\begin{figure}
\centerline{\psfig{file=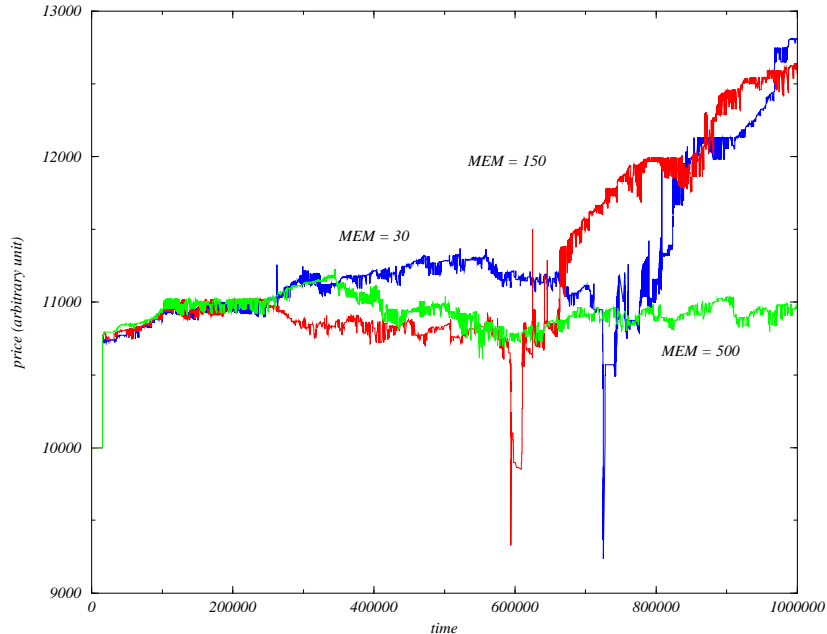,width=11cm,angle=270}}
\caption[]{\small\label{fig:memoria}
Influence of the length of the memory on simulation runs.}
\end{figure}

The effect of the parameter {\bf MEM} is shown in Fig.\ref{fig:memoria}. Large fluctuactions are present as a 
consequence of the herding behavior: Traders can ask each other about values and look at the history of the
market. But, as shown in Fig.\ref{fig:hurst}, correlations among prices are very short. As a consequence, if the
parameter {\bf MEM} becomes too big, then the agents tend to take into account events that are uncorrelated and
thus they perform a kind of averaging on the price: No trend has the change to fully develop itself because it is
only a small part of the window into which every trader is looking to.

\begin{figure}
\centerline{\psfig{file=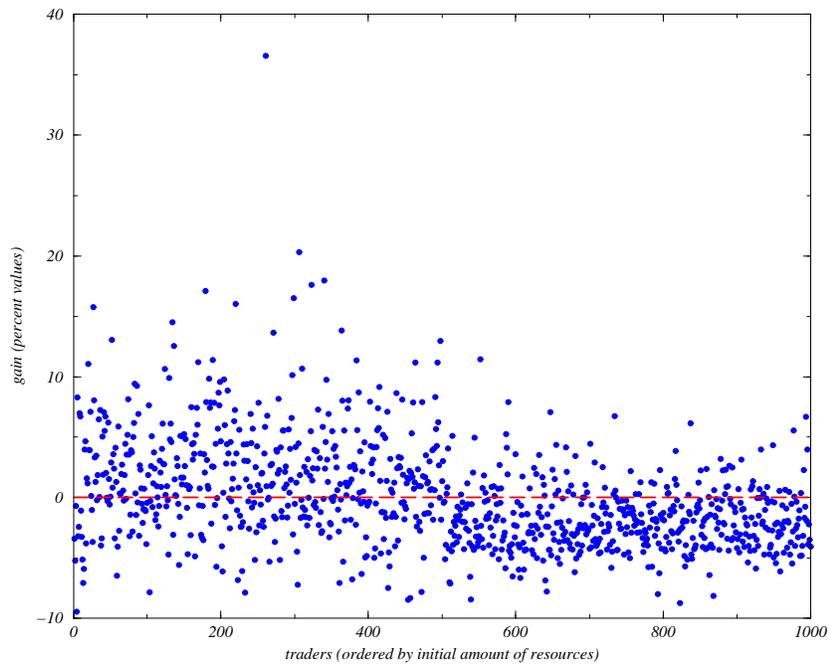,width=11cm,angle=270}}
\caption[]{\small\label{fig:vincite}
Distribution of the gain obtained by each trader. Agents are ordered according to the initial richness, with the richest
one in the leftmost position. Traders with less initial resources get systematically a worse performance, because
their actions are not able to affect significally the market.}
\end{figure}

In Fig.\ref{fig:vincite} we show the performance of each trader. Agents are ordered according to the
amount of money they have received at the beginning of the simulation: The leftmost trader was the richest one,
the rightmost was the poorest one. The gain is given in percent. We note a very interesting feature, namely traders
with more power (money) are systematically more lucky than the others. This is because they produce a bigger
echo when they trade: If one wants to buy a significant fraction of a stock, then the price will rise up and maybe
someone else will try to do the same, having identified a trend in the price and wanting to use this chance to make
money. As a consequence, the price may really increase of such an amount sufficient for the big trader to get the
desired profit and to decide to sell. After this operation, there is no immediate chance for the small agents to
get money, since the price is decreasing now as a consequence of the action of the powerful trader. Small agents
perform with a delay such that they are not able to get advantage from every speculative event. They are always 
waiting for the events, not being able to induce them.

\begin{figure}
\centerline{\psfig{file=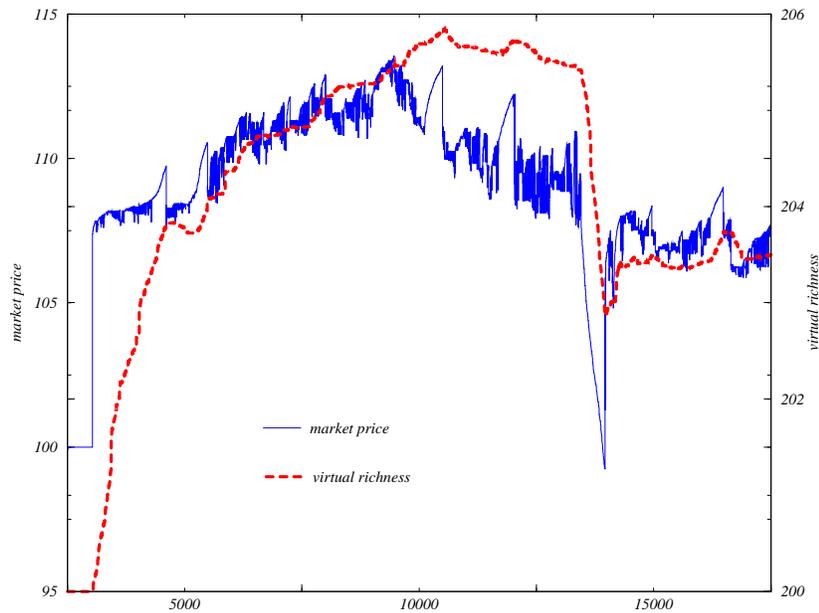,width=11cm,angle=270}}
\caption[]{\small\label{fig:ricchezza-virtuale}
On the dependance of the (virtual) richness on market price. This behavior is a consequence of the assumption that
the richness of a trader is given by the cash plus the invested money. In this way it seems that the closed market
is able to produce money.}
\end{figure}

Presenting this result, anyway, we assume the following, not so realistic, thing: The richness of a trader is given
by the sum of his/her cash and the money he/she has invested in shares. We should note that these two numbers are
different: One can use the latter provided he/she has sold the shares. Considering the invested money as
real money we suppose (without any reason) that the trader is able to sell the shares exactly at the buying price.
This is obviously not true. Fig.\ref{fig:ricchezza-virtuale} shows a consequence of this assumption: the virtual
richness increases (decreases) when the market price increases (decreases), with a small delay. We define virtual
this richness because one is sure to own a certain amount of share but not to have a certain amount of money, 
unless he/she is able to perform the conversion from shares to money just selling them.

The market we are simulating is a closed one, i.e. the total amount of involved money must remain 
constant\footnote{This would not be the case in presence of commissions, a percentage of money to be paid to the
bank for every transaction.}. But Fig.\ref{fig:ricchezza-virtuale} shows the opposite: The virtual richness is
positively correlated with the market price (delayed of a small amount of time). Although not shown in 
Fig.\ref{fig:ricchezza-virtuale}, if the price comes back to the IPO value\footnote{It should not only come back to the
IPO value, but also remain constant for a sufficiently long time, in order to let all the traders sell and buy the 
shares exactly at this price. In other words, it should happen that all the shares have been traded at the IPO price}, 
then the total richness comes back to the original value.
So, one way of getting rid of the problem is to replace the invested money with the number of owned shares times the 
IPO price. Doing so, the total richness remains constant, assuming the initially distributed value all the simulation 
long.

\begin{figure}
\centerline{\psfig{file=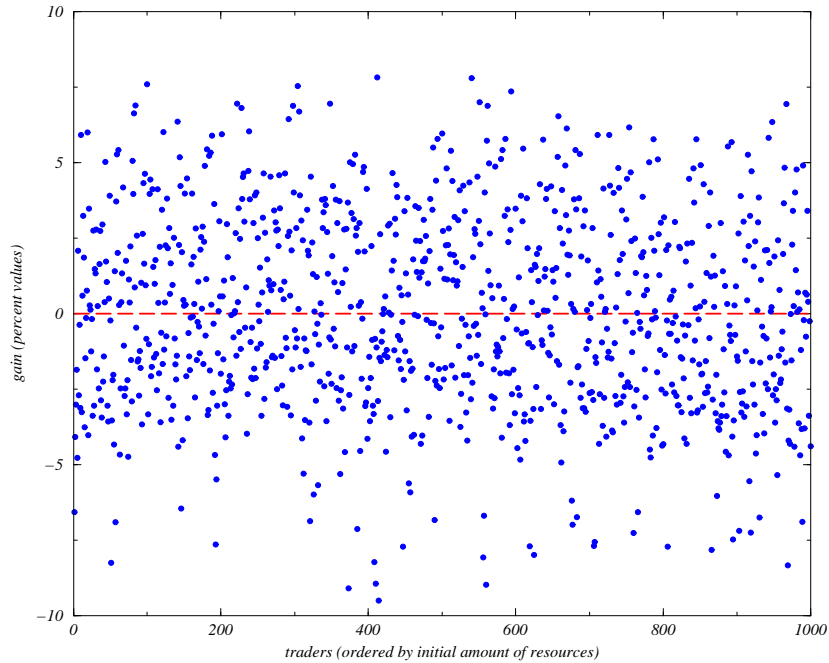,width=11cm,angle=270}}
\caption[]{\small\label{fig:servinc}
Distribution of the gain obtained by each trader. Agents are ordered according to the initial richness, with the richest
one in the leftmost position. The final richness is estimated as the sum of the cash and the
number of owned stocks time the IPO price.}
\end{figure}

In Fig.\ref{fig:servinc} we report the richness of every agent after the conversion from shares to cash,
with the IPO as buy-back price. The systematicity of Fig.\ref{fig:vincite} is no longer present. 
The reason is the following: Big traders loose all their advantages because just during a speculative period 
they cannot proceed and convert the shares into cash because we have decided to stop the simulation (and to 
convert their shares into cash at the IPO price instead of the current \emph{virtual} market price).

Maybe this is not a proper manner to act, but it is the only way to maintain isolated the system and to avoid
any artificial production of money. As a consequence of this abrupt end, many traders realize to have done a mistake
in performing a transaction because we impose from outside a price for the shares which is different compared to
the self organized one. This explains why according to the distribution of Fig.\ref{fig:servinc} all the traders
get similar performance independently from their initial status.

We are interested now in the association of a real temporal scale to our tick time. To this end, we have
performed a correlation analysis on our simulations and compared it to the results presented in \cite{mantegna,liu}.
Since the autocorrelation function vanishes after approximately 20 ticks and the typical correlation length in financial 
time series is supposed to be around 20 minutes, we can reach the conclusion that one of our tick corresponds to one 
real-life minute. This is the reason according to which we have decided to use the following value of the time related
parameters during the simulation runs:

\begin{itemize}

\item Total number of iterations, namely total number of ticks = $10^6$ ({\bf 10 years}).

\item Threshold = 10000 ({\bf 1 month}).

\item Threshold variability $[0.1,1]$. Combined with the threshold, this gives a time variable from a
      minimum of {\bf 1 month} to a maximum of {\bf 1 year}. After this time, traders start to ask themselves what
	  to do with the shares if the price remains too stable.

\end{itemize}

\section*{Conclusions}

In this paper we have introduced a model for the stockmarket whose parameters have a physical meaning. They have been
tuned as a direct consequence of observations of a typical trading activity. The main issue we wanted to address is that
our model is able to reproduce the two main characteristics of empirical data, namely correlated volatility and fat 
tails of the PDF of returns, with only one class of agents sharing the same goal: They all want to become as rich as
possible. In particular we do not need any external input nor a fundamental price, whose evolution would require some
artificial assumptions.
What we need is just a proper set of constraints: Traders have a limited amount of money, they can wait only for a
limited time, they have some desired gain in mind and they do not want to lose too much money. Furthermore they do not
invest all the money they have at disposal, since they want to use it in the most appropriate and less risky way.
We have proved the crucialilty of the parameters, showing the dependance of a typical simulation run from their
realistic tuning. Interpretations of the meaning of these values are also given, as they are mainly in agreement
with every-day situations.
In our opinion a model is really useful only as far as its parameter are completely understandable and influence directly
the evolution of the market price.

\end{document}